\documentclass[12pt]{book}
\usepackage{plenum}

\begin{document}

\chapter{NEW DEVELOPMENTS IN THE CONTINUOUS RENORMALIZATION GROUP}
\author{Tim R. Morris}

\affiliation{Department of Physics\\
University of Southampton\\
Highfield\\
Southampton, SO17 1BJ\\
U.K.}

\section{INTRODUCTION}

Over the last several years, there has been a 
resurgence of
interest in using non-perturbative approximation methods 
based on Wilson's continuous Renormalization
Group (RG), 
in quantum field theory.\refnote{\cite{kogwil}} 
The reason is simple: on the one hand
there is a desperate need for better analytic approximation methods to
understand truly non-perturbative situations in quantum field theory
(i.e. where there are no small parameters in which one can fruitfully
expand). On the other hand, Wilson's framework offers many possibilities
for systematic approximations which preserve a crucial defining property
of a quantum  field theory -- namely the existence of a continuum
limit.\refnote{\cite{erg},\cite{revi},\cite{revii}}

In this lecture, I review progress in the use and understanding of
Wilson's continuous renormalization group\refnote{\cite{kogwil}} 
particularly in
the past year. I try not to overlap too much with reviews given in
RG96\refnote{\cite{revii},\cite{yuri}}.
I concentrate on progress in the understanding of the structure of the
continuous RG since this is of fundamental importance to all research in
this area, and is an aspect that I have been particularly 
involved with, but I hope that the references collected at the end
are a representative  list
of just last years research in this area.\refnote{\cite{c1}--\cite{c11}} 
These papers deal with -- amongst other issues -- 
chiral symmetry breaking,\refnote{\cite{c1},\cite{c13}}
 chiral anomalies,\refnote{\cite{c2}}
finite temperature,\refnote{\cite{c13},\cite{c37},\cite{c37p},\cite{c3}}
 reparametrization invariance,\refnote{\cite{c4},\cite{com}}
gauge invariance,\refnote{\cite{c1},\cite{c5}}
perturbation theory,\refnote{\cite{c2},\cite{bonsim},\cite{c6}}
 gravity and supergravity,\refnote{\cite{c5}}
phase transitions,\refnote{\cite{c13},\cite{c37p},\cite{c4},
\cite{com},\cite{c7},\cite{mass}}
 novel
continuum limits,\refnote{\cite{c8}}
massive continuum limits,\refnote{\cite{mass}}
local potential approximation,\refnote{\cite{c9},\cite{AokN},\cite{cfn}}
large $N$ limits,\refnote{\cite{c9},\cite{AokN},\cite{lN},\cite{cfn}} 
$c$ functions,\refnote{\cite{cfn},\cite{c10}}
 and with applications from
condensed matter\refnote{\cite{c7}} to cosmology\refnote{\cite{c11}}.

The basic idea behind the (continuous) Wilsonian RG, is to construct the
partition function in two steps. Rather than integrate over all momentum
modes $q$ in one go, one first 
integrates out 
modes  between a cutoff scale $\Lambda_0$ and a very much
lower energy scale $\Lambda$.
Both of these scales are introduced by hand. 
The remaining integral from $\Lambda$ to zero may again be expressed as
a partition function, but the bare action $S_{\Lambda_0}$ (which is
typically chosen to be as simple a functional as possible) is replaced
by a very complicated effective action $S^{tot}_{\Lambda}$, and the
overall cutoff $\Lambda_0$ by the effective cutoff $\Lambda$.
Differential RG flow equations of the generic form 
\begin{equation}
\label{flo}
\Lambda{\partial\over\partial\Lambda} S^{tot}_\Lambda[\varphi] 
= {\cal F}[S^{tot}_\Lambda]\quad,
\end{equation}
can be written down that determine $S^{tot}_{\Lambda}$ in such a way 
that the resulting effective partition function gives precisely the
same results for all correlators as the original partition function. 

It may seem at first sight that such a partial integration step merely
complicates the issue. For example, we have had to replace the (generally)
simple $S_{\Lambda_0}$ by a complicated $S^{tot}_{\Lambda}$. However,
for the most part the complicated nature of $S^{tot}$ merely expresses
the fact that quantum field theory itself
is complicated: on setting $\Lambda=0$,
$S^{tot}_{\Lambda}$ becomes equivalent to the generating function for
all connected Green functions. 
To see this, note that the effective cutoff
$\Lambda$ can be regarded, either as an effective ultraviolet cutoff for
the remaining modes $q$ -- as just described, or, from the point of view of
the modes $q$ that have already been integrated out, $\Lambda$ behaves
as an infrared cutoff. These intuitive statements can be formalised
and proved\refnote{\cite{erg}}.\footnote{Somewhat similar statements 
have appeared elsewhere\refnote{\cite{oerg},\cite{wet}}}
  We introduce $\Lambda$ by modifying
propagators $\sim 1/q^2$ to $\Delta_{UV}=
C_{UV}(q,\Lambda)/q^2$, where $C_{UV}$
is a profile that acts as an ultra-violet cutoff\refnote{\cite{pol}},
i.e. $C_{UV}(0,\Lambda)=1$ and 
$C_{UV}\to0$ (sufficiently fast) as $q\to\infty$. If we introduce
the interaction part of the effective action as 
\begin{equation}
S^{tot}_\Lambda[\varphi] = {1\over2}\varphi\cdot{\Delta_{UV}^{-1}}\cdot\varphi
+ S_{\Lambda}[\varphi]\quad,
\label{Sdef}
\end{equation}
then Polchinski's form\refnote{\cite{pol}} of Wilson's 
RG\refnote{\cite{kogwil}} is,
\begin{equation}
{\partial S_{\Lambda}\over\partial\Lambda}
={1\over2}{\delta S_{\Lambda}\over\delta\varphi}\cdot{\partial\Delta_{UV}\over
\partial\Lambda}\cdot{\delta S_{\Lambda}\over\delta\varphi}-{1\over2}{\rm tr}
{\partial\Delta_{UV}\over\partial\Lambda}\cdot{\delta^2 S_{\Lambda}\over
\delta\varphi\delta\varphi}\quad.
\label{Pfl}
\end{equation}
On the other hand, $\exp -S^{tot}_{\Lambda}$ is itself
given by a partition function
$Z_{\Lambda}$, 
in which the effective cutoff appears as an infrared cutoff
as I have already mentioned.\refnote{\cite{erg}}
Propagators $\sim 1/q^2$ are in this partition function 
replaced by  $\Delta_{IR}=
C_{IR}(q,\Lambda)/q^2$, where it can be shown that
$C_{IR}(q,\Lambda)=1-C_{UV}(q,\Lambda)$. It is easy to see from the above
properties of $C_{UV}$, that $C_{IR}$ indeed behaves as an infrared cutoff.

From this partition function, $Z_\Lambda$,
 one can construct the Legendre effective
action $\Gamma_\Lambda^{tot}[\varphi^c]$, and it is immediately clear that
there must be a close relation between the two effective actions
$\Gamma^{tot}$ and $S^{tot}$. Indeed,
if we write the interaction parts of $\Gamma_\Lambda^{tot}$ as
\begin{equation}
\Gamma^{tot}_{\Lambda}[\varphi^c]={1\over2}\varphi^c\cdot\Delta_{IR}^{-1}\cdot\varphi^c
+\Gamma_\Lambda[\varphi^c]\quad,
\end{equation}
(where $\varphi^c$ is the so-called `classical' field) 
then it can be shown that the following Legendre 
transform type relation exists between the two effective 
actions\refnote{\cite{erg}}
\begin{equation}
S_\Lambda[\varphi]=\Gamma_\Lambda[\varphi^c]+{1\over2}(\varphi^c-\varphi)\cdot
\Delta^{-1}_{IR}\cdot(\varphi^c-\varphi)\quad.
\end{equation}
This relationship is important, for it means that the $\Lambda\to0$
limit of the Wilsonian effective action can be related to the standard
Legendre effective action $\Gamma[\varphi^c]=\lim_{\Lambda\to0}
\Gamma_\Lambda[\varphi^c]$ and hence to Green functions, S matrices,
classical effective potentials, and so forth. On the other hand,
by relating the infrared cutoff Legendre effective action to the
Wilsonian effective action, it provides physical justification
for the existence of fixed points and self-similar behaviour in
the solutions for $\Gamma_\Lambda$.
By differentiating $Z_\Lambda$, one readily obtains a flow equation for
the infrared cutoff Legendre effective 
action:\refnote{\cite{erg},\cite{deriv},\cite{wet},\cite{legfl}}
\begin{equation}
\label{Gfl}
{\partial\Gamma_\Lambda\over\partial\Lambda}=
-{1\over2}{\rm tr}\left[{1\over\Delta_{IR}}{\partial\Delta_{IR}\over\partial
\Lambda}\cdot\left(1+\Delta_{IR}\cdot{\delta^2\Gamma_\Lambda\over
\delta\varphi^c\delta\varphi^c}\right)^{-1}\right]\quad.
\end{equation}
It is straightforward to show
that substitution of the Legendre transform
relationship converts one flow equation into the other.

One should note also that Wilson's original form 
of the continuous RG\refnote{\cite{kogwil}} is related to Polchinski's
by some simple changes of variables\refnote{\cite{deriv},\cite{revii}},
and that the Wegner-Houghton sharp cutoff RG\refnote{\cite{wegho}} is nothing
but the sharp cutoff limit of Polchinski's 
equation\refnote{\cite{erg},\cite{truncm}}. Since these cover all present
forms of the continuous RG, this might give the impression that no other
forms are possible. This is not the case. But they are the simplest forms
of continuous RG (related to the fact that the cutoff $\Lambda$ may be
placed entirely within the inverse propagator). Other more complex versions
may eventually prove more useful for certain applications, e.g. gauge theory.

\section{STRUCTURE OF THE CONTINUUM LIMIT -- I}

This RG method of describing quantum field theory becomes advantageous when
we consider the continuum limit.  
I will indicate how  one can solve the flow equations in this case,
directly in the continuum, dispensing with the standard, but for quantum 
field theory, actually artificial and extraneous, scaffolding of imposing
an overall cutoff $\Lambda_0$, finding a sufficiently general bare action
$S_{\Lambda_0}$, and then tuning to a continuum limit as $\Lambda_0\to\infty$.
The solutions for the effective actions, being sensitive only
to momenta of magnitude $\sim\Lambda$, may be expressed
{\sl directly}
 in terms of renormalized quantities. In this case one finds
that the effective action may be expressed as a self-similar flow
of the relevant and marginally-relevant
 couplings, say $g^1$ to $g^n$, about
some fixed point: 
\begin{equation}
\label{ssfl}
S_\Lambda[\varphi]=S[\varphi]\left(g^1(\Lambda),\cdots,g^n(\varphi)\right)\quad.
\end{equation}
Actually, we also require to change to renormalised fields of 
course,\refnote{\cite{mass}} but, for clarities sake
 we leave this implicit. Also, to see this self-similar behaviour,
it is necessary to add the other essential ingredient of an RG step:
scaling back the cutoff to the original size. Simpler and equivalent, is
to ensure that all variables are `measured' in units of $\Lambda$, 
i.e. we change variables to
ones that are dimensionless, by
dividing by $\Lambda$ raised to the power
 of their scaling dimensions.\refnote{\cite{revii}}
From now on, we will assume that this has been done.

A fixed point $S_\Lambda[\varphi]=S_*[\varphi]$, i.e.
 such that $\partial S_*/\partial
\Lambda=0$,  being thus completely scale free,
corresponds to a massless continuum limit. We can arrange that the
coupling constants by definition vanish at the corresponding fixed point
\begin{equation}
\label{FP}
S[\varphi](0,\cdots,0)=S_*[\varphi]\quad,
\end{equation}
and are conjugate to the eigen-perturbations (i.e. integrated
operators of definite scaling dimension) at the fixed point:
\begin{equation}
{\partial S\over\partial g^i}\Big|_{g=0}= {\cal O}_i[\varphi]\quad.
\label{Os}
\end{equation}
We can easily
isolate a set of non-perturbative beta-functions\refnote{\cite{mass}}
$\beta^i(g)\equiv\Lambda\partial g^i/\partial
\Lambda$, whose perturbative expansion begins as
$\beta^i(g)=\lambda_i g^i +O(g^2)$, 
once a definition (a.k.a. renormalization)
condition consistent with (\ref{Os}) is chosen for the couplings.
Here, the $\lambda_i\ge0$ are the 
corresponding RG eigenvalues. This follows simply
by comparing coefficients of the left and right hand sides of (\ref{flo}).
Suppose for example, that ${\cal O}_4=\varphi^4$, and
a coupling $g^4$ is defined to be $c-c_*$
where $c(\Lambda)$ is the coefficient of the four-point vertex at zero
external momentum, 
and $c_*$ its
fixed point value; then the zero-momentum part of the four-point vertex on
the left hand side of (\ref{flo}) is nothing but $\beta^4$, and  the 
corresponding term on the right hand side yields an expression for it
in terms of $S_\Lambda$ itself. In this way one readily converts the
original Wilsonian RG into a self-similar flow for the underlying
relevant couplings, in close analogy to the usual field theory 
perturbative RG,\footnote{with the important difference however that
physical Green functions are here obtained
only when the scale used to define the beta function, 
tends to zero}
 although here the $\beta$ functions are defined
non-perturbatively.
Finite massive continuum limits then follow providing
that the couplings $g^i(\Lambda)$ are themselves finite 
at some finite physical scale $\Lambda\sim\mu$.\refnote{\cite{mass}}

Of course this translation to `manifest' self-similar flow is just a
rewriting, if the solution $S_\Lambda$ is already known. On the
other hand the solution $S_\Lambda$ {\sl is} determined completely
 (e.g. numerically) 
from (\ref{flo}), once 
the fixed point solution $S_*$ and relevant (and 
marginally-relevant) eigenperturbations are determined, since these
provide the complete boundary conditions via
\begin{equation}
\label{RTbc}
S_\Lambda=S_*+\sum_i \alpha^i (\mu/\Lambda)^{\lambda_i}
{\cal O}_i\quad\quad\quad
{\rm as} \quad \Lambda\to\infty\quad,
\end{equation}
where the $\alpha^i$ are the finite integration constants. (\ref{RTbc}) follows
from (\ref{FP}), (\ref{Os}) and the $\beta^i$ to first order in $g$.
Note that this more subtle boundary condition is required because 
$S_\Lambda=S_*$ is a singular point for the continuous RG, regarded as a  
first-order-in-$\Lambda$ differential equation.
In Wilsonian terms this establishes the initial position and direction
of the Renormalised Trajectory, which thus is sufficient to determine
the entire trajectory. In this form, or even better in
the self similar form of (\ref{FP}), (\ref{Os}) and the $\beta^i(g)$,
we have dispensed  entirely with the usual tuning procedure
required to reach a continuum limit. (We still need to compute the
infrared limit $\Lambda\to0$, however this involves no tuning,
and the asymptotic behaviour in this limit is straightforward
to derive analytically since it corresponds to `freeze-out' of
finite dimensionful quantities 
on the scale of $\Lambda<\!\!\!<\mu$.\refnote{\cite{mass}})
It is worth remarking also, that for a Gaussian fixed point these
equations (including the $\beta^i$) are 
 soluble analytically by iteration,
directly in terms of renormalised 
perturbation theory.\refnote{\cite{erg},\cite{cfn},\cite{bonsim}}

The remaining basic structural question  then, is
to understand how the (generically) finite number of
fixed point solutions and relevant perturbations arise from equations
such as (\ref{Pfl}) and (\ref{Gfl}), when these quantities are 
determined by functional differential equations (the right hand
sides of (\ref{Pfl}) or (\ref{Gfl}), and their perturbations) with
apparently
a continuous infinity of solutions. Let us return to this
after first considering some
possible approximations. Then  the so-called
Local Potential Approximation will be used 
to illustrate  the
general solution to these questions.\refnote{\cite{mass}}

\section{APPROXIMATIONS}

Clearly, except in very special simple cases, these flow equations are not
exactly soluble. However, virtually any approximation of the flow equations
that preserves the fact that they are non-linear (in $S_\Lambda$) will
continue to have fixed points and self-similar asymptotic solutions about
the fixed point of the form (\ref{ssfl}), i.e. preserve renormalisability.
This is in direct contrast to other frameworks for approximations such as
the use of Dyson-Schwinger equations.\refnote{\cite{erg}} The difficulty
then, is `only' one of finding approximations that are sufficiently
reliable and accurate for the purpose at hand.

The simplest form of approximation is to truncate the effective action
$S_\Lambda$ so that it contains just a few operators. The $\Lambda$
dependent coefficients of these operators then
have flow equations determined by equating coefficients on the left
and right hand side of (\ref{Pfl}), after rejecting from the right
hand side of (\ref{Pfl}) all terms that do not `fit' into this set
(this being the approximation). The difficulty with this approximation
is that it inevitably
results\footnote{unlike the case with truncations in the
real space renormalization group of simple spin systems, where such a
procedure has proved very succesful\refnote{\cite{revii}}} in a truncated
expansion in powers of the field $\varphi$ (about some point),
 which can only be sensible if
the field $\varphi$ does not fluctuate very much, i.e. is close to mean
field.\refnote{\cite{revii}} This is precisely the opposite regime
from the truly non-perturbative one that concerns us here. 
In most
situations, it is difficult in practice to be systematic about the choice
of which operators to include, while even if this is possible, 
in this regime one
finds generically that higher
orders cease to converge and thus yield limited accuracy, while there is
also no reliability -- even qualitatively -- since many spurious fixed
points are generated.\refnote{\cite{trunc}}

One exception to this rule deserves comment. As is well known,
fluctuations
of $\rho=\varphi^a\varphi_a$ 
(when appropriately scaled) disappear in  the limit $N\to\infty$. 
(Here $\varphi^a$ is an $N$ component scalar
field in an $O(N)$ invariant theory for example. 
More general large $N$ soluble systems
also have this property.) 
In this limit, it can be shown that the flow equations for 
the (infinitely many) operators
in $\Gamma_\Lambda$ which can be written in such a way that they only
involve $\rho$, form a closed set, so that truncation to all
such operators involves no approximation.\refnote{\cite{lN}}
Furthermore specialization
to just the 
non-derivative operators (thus forming the effective potential)
also involves no 
approximation.\refnote{\cite{wegho},\cite{c9},\cite{AokN},\cite{lN}}
In fact, the further truncation of these flow equations 
to a {\sl finite} power expansion in $\rho$ about this minimum, is
{\sl still} exact in this limit.\refnote{\cite{AokN}} This gives some
insight into why these latter truncations give
good results also
at finite $N$,\refnote{\cite{AokN},\cite{tetwet},\cite{goodtr},\cite{lN}}
improving at high orders of truncation to
 as much as 8 digits accuracy, before succumbing to
the generic  pattern 
for finite truncations  as 
outlined above.\refnote{\cite{Aokc},\cite{trunc}}

A less severe, and more natural, expansion, closely allied to the 
succesful truncations in real space RG of 
spin systems,\refnote{\cite{revii}} is rather to perform a `short distance
expansion'\refnote{\cite{revii},\cite{truncm}}
 of the effective action $\Gamma_\Lambda$, which for smooth cutoff
profiles corresponds to a derivative expansion.\refnote{\cite{deriv}} 
As well as evidently always allowing a systematic expansion (each level of
approximation merely corresponding to discarding all terms with more
than a specified number of derivatives), and 
robust and reasonably accurate approximations,\refnote{\cite{revii}}
it also preserves enough of the continuous RG to address the 
structural question posed above\footnote{in contrast to
the case of truncations to a finite set of operators}
 -- namely how (generically)
discrete sets of  fixed point solutions and eigenperturbations arise.
This is because the second order functional differential equations
that determine these,
reduce under the derivative expansion to second order ordinary
differential equations\refnote{\cite{deriv},\cite{twod},\cite{revii}}
which thus retain the property that they have a continuum of solutions.

The simplest such approximation is the so-called 
Local Potential Approximation (LPA),
introduced by Nicoll, Chang and Stanley:\refnote{\cite{nico}}
\begin{equation}
\label{LPA}
S_\Lambda\sim\int\!\!d^D\!x\,\left\{{1\over2}
(\partial_\mu\varphi)^2+V(\varphi,\Lambda)
\right\}\quad.
\end{equation}
It has since
been rediscovered by  
many authors,\refnote{\cite{truncm}}
notably Hasenfratz and Hasenfratz.\refnote{\cite{hashas}}
As a concrete example, consider the case of sharp cutoff. The flow equations
may be shown to reduce to\refnote{\cite{nico},\cite{hashas},\cite{erg},
\cite{trunc},\cite{truncm}}
\begin{equation}
\label{floV}
{\partial\over\partial t}V(\varphi,t)+{1\over2}(D-2)\,\varphi V'
-DV=\ln(1+V'')\quad,
\end{equation}
where $\prime\equiv{\partial\over\partial\varphi}$ and
$t=\ln(\mu/\Lambda)$. 
Actually, the analogous $N=\infty$ case of (\ref{floV}) 
was already derived by Wegner and
Houghton in their paper introducing the sharp-cutoff flow 
equation.\refnote{\cite{wegho}}
As already pointed out above, in this limit the LPA is effectively
exact.\refnote{\cite{lN}} It can be 
shown\refnote{\cite{lN},\cite{mass},\cite{erg}} that $V$ coincides with the
Legendre effective potential in the limit $\Lambda\to0$, i.e. $t\to\infty$.

\section{STRUCTURE OF THE CONTINUUM LIMIT -- II}

Thus in this example, the fixed point potential satisfies
\begin{equation}
\label{V*}
{1\over2}(D-2)\,\varphi V'_*(\varphi)
-DV_*(\varphi)=\ln(1+V''_*)\quad.
\end{equation}
This equation has indeed a continuum of solutions, in fact a 
continuous two-parameter set. However generically,
all but a countable number
of these solutions are singular!\refnote{\cite{trunc},\cite{hh}}
 ($D=2$ dimensions is an exception.\refnote{\cite{twod}})
It is the discrete set of non-singular solutions that are
approximations to the exact fixed points.
A review was given in RG96.\refnote{\cite{revii},\cite{hh}}
Since this scenario generalises to higher orders of the derivative
expansion (with an increasingly larger dimension
parameter space of solutions),\refnote{\cite{deriv},\cite{revii}}
 it seems reasonable to suppose that it applies also to the exact
RG equations (e.g. \ref{Pfl},\ref{Gfl}) in the sense that, although
the second order non-linear functional differential equations 
governing the fixed points have a full functional space worth of
solutions, only typically a discrete number are well defined
for all $\varphi({\bf x})$.\refnote{\cite{mass}}

As reviewed in RG96,\refnote{\cite{revii}} 
this structure
of the derivative expansion equations, and (\ref{V*}), allows 
straightforwardly for systematic
searches of all possible continuum limits  within
this approximation. In the traditional approaches, this would require
systematic searches of 
the infinite dimensional space of all possible bare actions!

For large field $\varphi$ the only consistent behaviour (with $D>2$) for
the fixed point potential in (\ref{V*}) is 
\begin{equation}
\label{V*as}
V_*(\varphi)\sim A \varphi^{D/d}\quad,
\end{equation}
where $d={1\over2}(D-2)$, and $A$ is a constant determined by
the equations. 
This simply solves the left hand side of (\ref{V*}), 
these terms arising from purely dimensional considerations,
and neglects
the right hand side of the flow equation -- which encodes the
quantum corrections. Or in other words,
(\ref{V*as}) is precisely what would be expected by dimensions
(since $V$'s mass-dimension is $D$ and $\varphi$'s is $d$)
providing only that any dependence on $\Lambda$, and thus
the remaining quantum corrections, can
be neglected. Requiring the form (\ref{V*as}) to hold for both
$\varphi\to\infty$ and $\varphi\to-\infty$, provides 
the necessary two boundary
conditions for the second order ordinary differential equation (\ref{V*}),
so we should indeed generally
expect at most a discrete set of globally non-singular solutions.
These considerations generalise to any order of the derivative expansion,
and indeed we expect them to hold also for the exact RG, with one
modification: beyond LPA, $d={1\over2}(D-2+\eta)$, where $\eta$ is
the anomalous dimension at the fixed point.\refnote{\cite{mass}}
 
($\eta$ is set artificially
to zero by the LPA because in the LPA the momentum dependent terms
in (\ref{LPA}) remain uncorrected. There are subtleties to
do with reparametrization invariance, if higher orders in the
 derivative expansion
are to determine a discrete set of solutions for $\eta$, as properly
to be expected.
We will not review them again here. The preservation of such an
invariance under the
derivative expansion, requires a particular power-law form of cutoff 
profile.\refnote{\cite{revii},\cite{revi},\cite{deriv}}
There has been recent progress in understanding the ramifications 
of reparametrization invariance for
derivative expansions of the Polchinski equation.\refnote{\cite{com}})

Now consider the determination of the eigenoperators. For this
we set (by separation of variables),
\begin{equation}
\label{VRGbc}
V(\varphi,t)=V_*(\varphi)+\alpha {\rm e}^{\lambda t} v(\varphi)\quad,
\end{equation}
and expand to first order in $\alpha$ (c.f. (\ref{RTbc})),
obtaining
\begin{equation}
\label{v}
\lambda v + d\,\varphi v'-Dv={v''\over1+V''_*}\quad.
\end{equation}
In this case we again have a one parameter continuum of solutions
(after some choice of normalisation), which are guaranteed globally
well defined since (\ref{v}) is linear,
and this for each value of $\lambda$. How can this be squared with the
fact that experiment, simulation etc., typically only uncover a 
discrete spectrum of such operators? The answer is that only the
discrete set of solutions for $v(\varphi)$ that behave as a power of
$\varphi$ for large field, can be associated with a corresponding 
renormalised coupling $g(t)$ and thus the 
universal self-similar flow (\ref{ssfl})
which is characteristic of
 the continuum limit.\refnote{\cite{hh},\cite{mass}}

Indeed we see from (\ref{v}) and (\ref{V*as}), that those solutions
that behave as a power for large $\varphi$ must do so as
\begin{equation}
\label{vpo}
v(\varphi)\sim  \varphi^{(D-\lambda)/d}\quad,
\end{equation}
this being again the required power to balance scaling dimensions 
(with $[g(t)]=\lambda$) if the remaining quantum corrections may
be neglected in this regime. Once again for $\varphi\to\pm\infty$, this
supplies two boundary conditions, but 
this time, since (\ref{v}) is linear, this
overdetermines the equations, and generically
allows only certain quantized values of $\lambda$. 

On the other hand if $v$ does not behave as a power of $\varphi$ for
large $\varphi$, then from (\ref{v}) and (\ref{V*as}), we obtain
that instead
\begin{equation}
\label{vnonpo}
v(\varphi)\sim\exp\left\{A(D-d)\varphi^{D/d}\right\}\quad.
\end{equation}

Actually, in the large $\varphi$ regime, we may solve (\ref{floV})
without linearising in $\alpha$. Indeed we may solve it 
non-perturbatively. This is because, just as before,
we may neglect the quantum corrections in this regime.
These are given by the right hand side of (\ref{floV}),
and thus $V(\varphi,t)$ follows 
mean-field-like evolution:
\begin{equation}
\label{mf}
V(\varphi,t)\sim {\rm e}^{Dt} V(\varphi{\rm e}^{-dt},0)\quad.
\end{equation}
If we take $V(\varphi,0)$ to be given by (\ref{VRGbc}) at $t=0$, 
and use (\ref{mf}), (\ref{V*as}) and (\ref{vpo}), then we 
see that for large $\varphi$, we recover the $t$ dependence of (\ref{VRGbc})
even without linearising in $\alpha$. Thus for power law $v$ (\ref{vpo})
in the large $\varphi$ regime,
we may absorb the $t$ dependence into the self-similar flow
of a corresponding coupling $g(t)=\alpha {\rm e}^{\lambda t}$,
even for finite $\alpha$.
But using (\ref{mf}) and
the non-power-law behaviour (\ref{vnonpo}), results in the
$t$ dependence being `stuck' in the exponential in (\ref{vnonpo}).
In this case then, the scale dependence cannot be combined with $\alpha$
into a corresponding coupling and the RG flow is not self-similar.
In fact one can further show that perturbations of the
form (\ref{vnonpo}) collapse on $t$-evolution into an infinite
sum of the quantized power-law perturbations, and thus the non-power
law eigenperturbations are entirely irrelevant for 
continuum physics.\refnote{\cite{mass}}

Again, these considerations generalise to higher orders of the derivative
expansion and thus presumably to the exact RG, the non-quantized
perturbations 
 growing faster than a power for large
$\varphi$ and consequently not associated with renormalised 
couplings. (The
precise form of the large $\varphi$ dependence of the 
non-power-law perturbations
however depends on non-universal details including the level
of derivative expansion approximation used, if any.)\refnote{\cite{mass}}

I finish the review with a couple of applications of these ideas.

\section{A NUMERICAL EXAMPLE}

It is worth stating again that the derivative
expansion gives fair numerical approximations in 
practice.\refnote{\cite{revii}}
For example, we have previously reviewed\refnote{\cite{revii}}
 the numerical results for
$N$ component scalar field theory at the non-perturbative Wilson-Fisher
fixed point in three dimensions,\refnote{\cite{c4}}
and the impressive
numerical results for the sequence of multicritical scalar field
theories in two dimensions (where all standard 
methods fail).\refnote{\cite{twod}}
As an example of the application of all
the above concepts to a calculation,
we here review
the Ising model scaling equation of state, in the symmetric phase, in
three dimensions.\refnote{\cite{mass}} This is of current interest
 since it allows direct comparison with the recent progress in
accurate calculations from 
resummed perturbation theory.\refnote{\cite{guzin},\cite{sok},\cite{zakzin}}
The Ising model
equation of state follows
from the Legendre effective potential of the massive
non-perturbative field theory
of a single scalar field built around the Wilson-Fisher fixed point.
Such a theory has only one relevant eigenperturbation
and thus only one coupling -- which simply sets the scale. If all
quantities are  measured in terms of this mass-scale, the results
are thus pure numbers and universal. (This is of course the basis for the
universality of critical phenomena in the Ising model class.)

A second order derivative expansion, `$O(\partial^2)$',
\begin{equation}
\Gamma_t=\int\!\!d^D\!x\,\left\{{1\over2}(\partial_\mu\varphi)^2
K(\varphi,t)+V(\varphi,t)
\right\}
\end{equation}
with a certain careful choice of smooth cutoff ($\sim$ power-law as
alluded to earlier) results in the flow equations,\refnote{\cite{deriv}}
\begin{eqnarray}
&&\label{Vflod2}\phantom{\hbox{and}\hskip 1cm}
{\partial V\over\partial t}+{1\over2}(1+\eta)\varphi V'-3V=
-{1-\eta/4\over\sqrt{K}\sqrt{V''+2\sqrt{K}} } \\
&&\label{Kflo}\hbox{and}\hskip 1cm
{\partial K\over \partial t}+{1\over2}(1+\eta)\varphi K' +\eta K=
\left(1-{\eta\over4}\right)\Biggl\{ 
{1\over48}{24KK''-19(K')^2\over K^{3/2}(V''+2\sqrt{K})^{3/2}} \\
&&\nonumber
-{1\over48}{58V'''K'\sqrt{K}+57(K')^2+(V''')^2K\over K(V''+2\sqrt{K})^{5/2}}
+{5\over12}{(V''')^2K+2V'''K'\sqrt{K}+(K')^2\over\sqrt{K}(V''+2\sqrt{K})^{7/2}}
\Biggr\}\ \ .
\end{eqnarray}
These equations enjoy the following reparametrization symmetry,
$\varphi\mapsto\Omega^{5/2}\varphi$, $V\mapsto\Omega^3V$, $K\mapsto\Omega^{-4}K$,
which turns the fixed point equations into non-linear eigenvalue
equations for the anomalous dimension 
$\eta$.\refnote{\cite{deriv},\cite{revii}} 
 On the other hand, by discarding (\ref{Kflo}), and setting
$K=1$ and $\eta=0$
 in (\ref{Vflod2}), one obtains the flow equation associated with the
lowest order in the derivative expansion (with this cutoff), `$O(\partial^0)$'.

It should be pointed out that one method of extracting the universal
characteristics of the fixed point behaviour in the above equations
is simply to solve them in a traditional way, by choosing an
appropriate bare potential
and bare $K$ (typically $K=1$), at some initial point $t_0$ (effectively
the cutoff $\Lambda_0$) and tuning 
the potential as the cutoff is removed, 
so as to recover a continuum limit. 
Nevertheless, the numerical methods based on the insight of the previous
sections are certainly faster, more elegant, and more accurate.

We will not detail the numerical methods used to implement 
the above ideas,\refnote{\cite{c4},\cite{revii},\cite{twod},\cite{deriv}}
but simply quote the results.
We found just one non-trivial
non-singular fixed point solution with 
$\eta=.05393208\cdots$, which is thus identified with 
the Wilson-Fisher fixed point.\refnote{\cite{deriv},\cite{mass}}
Our resulting $O(\partial^2)$ value for $\eta$ 
should be compared with the (combined) worlds best 
estimates\refnote{\cite{zinn},\cite{deriv}}
$\eta=.035(3)$. Extracting from the quantized spectrum of power-law
eigenperturbations, the relevant operator
and the least irrelevant operator, one obtains the critical exponents
$\nu$ and $\omega$ (via the formulae\refnote{\cite{kogwil}}
$\nu=1/\lambda$ and $\omega=-\lambda$). We found at $O(\partial^0)$,
$\nu=.6604$, $\omega=.6285$, and at $O(\partial^2)$, $\nu=.6181$
and $\omega=.8972$. These should be compared to combined worlds
best estimates of $\nu=.631(2)$ 
and $\omega=.80(4)$.\refnote{\cite{zinn},\cite{deriv}}

Having transformed to renormalised variables and integrating out along
the renormalised trajectory we have that the Legendre effective potential
$V(\varphi)$ is given by the $t\to\infty$ limit of $V(\varphi,t)$. Writing in
terms of the physical dimensionful variables
\begin{equation}
V(\varphi)={1\over2}m^2\varphi^2+u_4\varphi^4+u_6\varphi^6+\cdots,
\end{equation}
the ratios $\alpha^{\rm p}_{2k}=u_{2k}m^{k-3}$ are dimensionless and universal.
By Taylor expansion of the equations (\ref{Vflod2},\ref{Kflo}), careful
choice of closure ansatze, and reworking the
equations to expose the self-similar flow along the renormalised
trajectory (in the way indicated earlier), 
we obtained ordinary differential equations which could be
straightforwardly integrated  to obtain the 
$\alpha^{\rm p}_{2k}$.\refnote{\cite{mass}}
(We performed all the calculations within
the Maple package.) The results are displayed in table 1.
The six-point coupling and higher are written
in terms of $F_{2k-1}=2k\alpha^{\rm p}_{2k}/
(24\alpha^{\rm p}_4)^{k-1}$, which allows a direct comparison with
the most accurate recent perturbative results.\refnote{\cite{guzin}}
In row ``Sharp'' of this table, we show also the results obtained
from eqn.(\ref{floV}). In the row ``$\partial$ exp$^{\rm n}$'', we 
use both orders of the derivative expansion, and a comparison of our
results
for the critical exponents $\omega$ and $\nu$ with the best determinations,
 to estimate an error.\refnote{\cite{mass}}
It should not be taken as seriously as
the very careful error analyses possible, and 
performed,\refnote{\cite{guzin}}
in the large order perturbation theory calculations.
Those rows labelled $D$=3, $\varepsilon{-\rm exp.}$, ERG, HT, and MC
give results respectively from resummed perturbation
theory, $\epsilon$ expansion, another  exact RG approximation, 
high temperature series and Monte-Carlo estimates.

\atable{Universal coupling constant ratios for the three dimensional 
Ising model universality class.}
{\begin{tabular}{ccccc}
\hline
 Approximation & $\alpha^{\rm p}_4$ & $F_5$ & $F_7$ & $F_9$  \\ \hline
Sharp &1.514(2) &.0155(3) &$3.6(5)\times10^{-4}$&
$-1.7(5)\times10^{-5}$ \\
$O(\partial^0)$&1.3012(2)& .01638(1) & $4.68(3)\times10^{-4}$& 
$-2.4(1)\times10^{-5}$ \\
$O(\partial^2)$& .8635(5)& .01719(4)& $4.9(1)\times10^{-4}$ 
& $-5.2(3)\times10^{-5}$  \\ \hline
$\partial$ exp$^{\rm n}$&.86(15)&.0172(3)&$4.9(1)\times10^{-4}$&
$-5(1)\times10^{-5}$ \\ \hline
$D$=3~\refnote{\cite{guzin}} &.988(2)&.01712(6)&$4.96(49)\times10^{-4}$&$
 -6(4)\times 10^{-5} $ \\
$D$=3~\refnote{\cite{sok}}
 & \hfil  & .0168 - .0173 &  4.1 - 16.2$\,\times10^{-4}$ & \\
$\varepsilon{-\rm exp.}$~\refnote{\cite{guzin}} & $1.2$& $ 
.0176(4) $&$  4.5(3)\times 10^{-4}  $&$
 -3.2(2) \times 10^{-5} $ \\
$\varepsilon{-\rm exp.}$~\refnote{\cite{sok}} &\hfil  & .0176 & & \\
 ERG~\refnote{\cite{tetwet},\cite{ber}}
 &$  1.20  $&$ .016 $&$  4.3\times 10^{-4} $&$   $ \\
HT~\refnote{\cite{rreisz}} & .99(6) & .0205(52)&$    $&$  $ \\
 HT~\refnote{\cite{rZLFish}} &$1.019(6)$&$ .01780(15)$&$   $&$   $ \\
 HT~\refnote{\cite{rbuco}} &.987(4)&.017(1)& $5.4(6)\times10^{-4}$ &
$-2(1)\times10^{-5}$ \\
 MC~\refnote{\cite{Tsyp}} &$  .97(2)  $&$ .0227(26) $&$   $&$   $ \\
 MC~\refnote{\cite{rkimlandau}}
 &$  1.020(8) $&$ .027(2) $&$  .00236(40)  $&  
\\ \hline
\end{tabular}}

It can be seen from the table,
that while perturbative methods are more powerful than
the derivative expansion for low order couplings, the derivative expansion
eventually wins out. The reason for this is that the derivative expansion
at these lowest orders, is crude in comparison to the perturbation
theory methods, however the perturbative methods suffer from being asymptotic
 -- which in particular results in rapidly worse determinations for
higher order couplings. The derivative expansion does not suffer from this,
since it is not related at all to an expansion in powers of the field.
Indeed, it may be shown that\refnote{\cite{erg}}
 even at the level of the LPA,
Feynman diagrams of all topologies are included.

Let us mention that we also obtained `for free'
some estimates for $F_{11}$ and $F_{13}$,
and universal coefficient ratios in the $O(\partial^2)$ function
$K(\varphi)$.\refnote{\cite{mass}}
 These latter correspond to universal information about the
$O(p^2)$ terms of the one particle irreducible Green functions. 
At present, we know
of no other estimates with which these can be compared.

\section{THEORETICAL EXAMPLE -- A C FUNCTION}

Following Zamolodchikov's celebrated 
$c$-theorem\refnote{\cite{zam}}
for two dimensional quantum field theory, a number of groups have sought
to generalise these ideas to higher 
dimensions.\refnote{\cite{c10},\cite{zamfan}} The
motivation behind this, is not only to demonstrate irreversibility
of renormalization group flows, 
and thus prove that exotic flows such as limit cycles,
chaos, etc, are missing in these cases, but perhaps more importantly to
provide an explicit, and useful, geometric framework for the space of
quantum field theories.
Zamolodchikov established three important properties for his
$c$ function:
\medskip

(i)\ \ There exists a function $c(g)\ge0$ of such a nature that
${d\over dt} c\equiv \beta^i(g){\partial\over\partial g^i} c(g)
\le 0$,

the equality being  reached only at fixed points  $g(t)=g_*$.

(ii)\ \ $c(g)$ is stationary at 
fixed points,
i.e. $\beta^i(g)=0$ for all $i$, implies $\partial c/\partial g^i=0$.

(iii)\ \ The value of $c(g)$ at the fixed point $g_*$ is the same as
the corresponding 

Virasoro algebra central charge.\refnote{\cite{cft}}
 (This property thus
only makes sense in two 

dimensions.)
\medskip

Within the LPA to Wilson's, or
Polchinski's exact RG,\footnote{for $N$ scalar fields,
with no symmetry implied, and (almost) any smooth 
cutoff.\refnote{\cite{cfn},\cite{ball}}}
\begin{equation}
\label{PLPA}
{\partial\over\partial t} V(\varphi,t) +{1\over2}(D-2)\varphi_a{\partial V\over
\partial\varphi_a} - D V = 
 {\partial^2V\over\partial\varphi_a\partial\varphi_a}
-\left({\partial V\over\partial\varphi_a}\right)^2
\end{equation}
we have discovered\refnote{\cite{cfn}} a
$c$-function which has the first two properties in any dimension $D$.
Our $c$-function has a counting property which
 generalises property (iii): it is extensive at fixed points, 
i.e. additive in
mutually non-interacting degrees of freedom,
and counts one for each Gaussian
scalar and zero for each infinitely massive scalar (corresponding to a
High Temperature fixed point). These properties are shared by the
two dimensional Virasoro central charge. 
It is probably not possible within the LPA, 
to establish a more concrete link to Zamolodchikov's $c$.

The idea is very simple and builds on the observation of 
Zumbach,\refnote{\cite{zum}} that (\ref{PLPA}) may be written as a gradient
flow
\begin{equation}
\label{Fflo}
a^N G {\partial \rho\over\partial t} =-{\delta{\cal F}\over\delta\rho}\quad,
\end{equation}
where we have introduced $G=\exp-{1\over4}(D-2)\varphi^2_a$, $\rho={\rm e}^{-V}$,
and a measure normalization factor $a$. The functional ${\cal F}$ is
given by
\begin{equation}
{\cal F}[\rho]=a^N\!\!\!\int\!\! d^N\!\varphi\ G\left\{
{1\over2}\left({\partial\rho\over\partial\varphi_a}\right)^2
+{D\over4}\rho^2\left(1-2\ln\rho\right)\right\} \quad.
\end{equation}
Fixed points $\rho=\rho_*$
are then given by $\delta{\cal F}/\delta\rho=0$.
If one substitutes this equation back into ${\cal F}$ one 
obtains\refnote{\cite{cfn}}
\begin{equation}
\label{fpF}
{\cal F}[\rho_*]={D\over4}\,a^N\!\!\!\int\!\! d^N\!\varphi\,G\rho^2_*\quad.
\end{equation}
Now, if the field content splits into two mutually non-interacting
sets, the potential splits in two, and thus $\rho_*$ factorizes.
We see then that the integral in (\ref{fpF}) itself factorizes.
Thus if we define a $c$ function through the logarithm of ${\cal F}$,
$c=\ln(4{\cal F}/D)/A$ (where $A$ is another normalization factor),
it will be additive in mutually non-interacting degrees of freedom.
 Furthermore, we may use the 
normalization constants $a$ and $A$, to normalise the $c$ function
so that it counts one for Gaussian scalars ($V_*=0$)
and zero for the high
temperature fixed point ($V_*={1\over2}\varphi^2-1/D$ per 
scalar\refnote{\cite{c9},\cite{cfn}}).

The other properties follow from the geometrization of (\ref{Fflo}), by
rewriting the flow in terms of a complete set of coupling constants $g^i(t)$,
$V\equiv V(\varphi,g)$.
Close to a fixed point, these need only be the relevant and marginal
couplings, as we discussed earlier. Elsewhere, they have to be infinite in
number to span the space of all potentials. Following 
Zamolodchikov\refnote{\cite{zam}} we generalise (\ref{Os}) to the whole
space by writing ${\cal O}_i(g)=\partial_i V(\varphi,g)$, where
$\partial_i\equiv\partial/\partial g^i$.
Then, after some straightforward manipulation, we obtain
\begin{equation}
\label{cflo}
\partial_i c(g) = -{\cal G}_{ij}\beta^j(g)\quad,
\end{equation}
where ${\cal G}_{ij}$ is our analogue of the so-called Zamolodchikov metric 
on coupling constant space:\refnote{\cite{zam}}
\begin{equation}
{\cal G}_{ij}(g)=({\cal F}\ln A)a^N\!\!\!\int\!\! d^N\!\varphi\,
G\rho^2\,{\cal O}_i{\cal O}_j\quad.
\end{equation}
Property (ii) follows immediately from (\ref{cflo}), while,
because ${\cal G}_{ij}$ is positive definite, multiplication of (\ref{cflo})
by $\beta^i$ establishes property (i).

I close with two remarks. Firstly, it would be very interesting of course,
to understand if this LPA $c$ function generalizes
to higher orders in the derivative expansion\refnote{\cite{twod}}
(which likely would allow a direct comparison with Zamolodchikov's
$c$), or indeed generalizes to an exact expression along the present lines.
Secondly, 
the geometric structure immediately 
suggests a variational approximation method for (\ref{PLPA}): 
Restricting the flows
to some finite dimensional submanifold parametrized 
(in some way) say by $g^1,\cdots,g^M$,
 corresponds to ansatzing a subspace of potentials $V\equiv
V(\varphi,g)$, where $g$ stands now only for the $M$ parameters. 
Approximations to the fixed points then
follow from the variational equations $\partial_ic(g)=0$.
Our investigations indicate that this approximation method
is rather powerful.\refnote{\cite{cfn},\cite{wip}}
This illustrates once again that purely
theoretical insights into the structure of 
the continuous RG and its approximations, 
tend in turn to suggest yet more elegant and more
powerful methods of approximation.

\end{document}